\documentclass[aps,prl,groupedaddress,amsmath,amssymb,floatfix,twocolumn]{revtex4-1}
\usepackage[utf8]{inputenc}
\usepackage{lmodern}
\usepackage[pdftex]{graphicx}
\usepackage{hyperref}
\usepackage{color}
\usepackage{multirow}
\usepackage{subcaption}
\usepackage[cal=esstix]{mathalfa}

\newcommand{\kay}{\mathcal{k}}
\newcommand{\GSC}{{\texttt{GSC} }}
\newcommand{\ILG}{{\texttt{ILG} }}

\begin{document}

\title{A Network of Networks Approach to Interconnected Power Grids}

\author{Paul~Schultz$^{1,2}$}
\email{pschultz@pik-potsdam.de}
\author{Frank~Hellmann$^{1}$}
\author{Jobst~Heitzig$^{1}$}
\author{J\"{u}rgen~Kurths$^{1,2,3,4}$}

\affiliation{$^1$Potsdam Institute for Climate Impact Research, P.O. Box 60 12 03, 14412 Potsdam, Germany\\
$^2$Department of Physics, Humboldt University of Berlin, Newtonstr. 15, 12489 Berlin, Germany\\
$^3$Institute for Complex Systems and Mathematical Biology, University of Aberdeen, Aberdeen AB24 3UE, United Kingdom\\
$^4$Department of Control Theory, Nizhny Novgorod State University, Gagarin Avenue 23, 606950 Nizhny Novgorod, Russia}

\begin{abstract}
We present two different approaches to model power grids as interconnected networks of networks.
Both models are derived from a model for spatially embedded mono-layer networks and are generalised
to handle an arbitrary number of network layers. The two approaches are distinguished by their use case.
The static glue stick construction model yields a multi-layer network from a predefined layer interconnection scheme, i.e. 
different layers are attached with transformer edges. It is especially suited to construct multi-layer
power grids with a specified number of nodes in and transformers between layers. We contrast it
with a genuine growth model which we label interconnected layer growth model.
\end{abstract}


\maketitle

\section{Introduction}
\label{sec:introduction}

The power grid has been described as the most complex machine ever built. 
It consists of millions of individual machines that interact through a 
hierarchical multi-layer network that scales from single buildings to continents. 
Currently most studies of the power grid focus on a particular layer. This was 
possible as in the past, very well-defined roles have been assigned to different 
layers, being still evident in their names. The transmission layer would transmit 
energy from big centralised producers to centres of consumption and demand, 
balancing the overall energy budget. The distribution network would then distribute that 
energy to the individual consumers through a further hierarchy of voltage levels.

This clear assignment of particular functions to voltage level increasingly no longer 
reflects the real structure of the power grid. As a consequence of, for instance, consumers 
operating solar home systems and smaller wind parks being connected at mid-voltage levels, 
distribution grids need to collect as well as aggregate energy. In the future, micro grids will try 
to achieve a more local balance of production and consumption, while also providing 
cross layer stability services that were not needed or possible previously.

So far distribution grids have been seen mostly as passive consumers of stability from the top down. 
Frequency stability was entirely in the remit of the transmission grid operators, while voltage stability 
in the two grid layers was considered a separate issue.
With distributed generation capacity in the distribution grids, a novel topic in power grid design is how 
stability services can be aggregated from the distribution grid into the transmission grid \cite{Auer2016,Marten2013}. 
This includes, for instance, the provision of reactive power as well as virtual swing mass to the transmission grid.
The grid topology plays a large role in whether the distributed ancillary services can be provided to the transmission grid. 
Few systematic results on this influence exist so far, so it is necessary to consider the impact on a case by case basis.

In the study of individual network layers it has been increasingly understood that it 
is necessary to study the properties of power grids not just for individual benchmark 
models but also for ensembles of plausible topologies. In this respect, the DENA ``Verteilnetzstudie'' \cite{dena2012} 
used several representative ensembles of distribution grids and work on the dynamic stability 
of transmission systems has recently been advanced considerably by the availability of 
random network models that capture many typical properties of transmission systems \cite{Schultz2014,Soltan2015}.

These random networks allow us to have sufficient statistics to identify particular 
topological features that can have a very large influence on non-linear dynamical properties 
like basin stability and survivability \cite{Menck2013,Menck2014,Schultz2014a,Plietzsch2015,Nitzbon2016}, 
but also resilience against cascading failures caused by single disruptive events
 \cite{Buldyrev2010,Shao2011,Brummitt2012,Zhao2016}.

In this paper we introduce a random network algorithm that generalises this approach and captures
the hierarchical multi-layer nature of a combined distribution and transmission grid. We contrast 
and compare two different mechanisms by which we can arrive at such a network, static construction
(top-down or bottom-up) or concurrent growth. 
Both of these algorithms have different applications. The static construction approach 
is able to model the power grids of existing industrialised countries that are introducing 
renewables now as well as networks that are now coming into existence 
in the developing world, and that are renewable and micro grid driven from the start. The 
growth algorithm, however, does not rely on input from real power grids like, e.g. the ratio
of intra- vs. inter-layer links for pairs of network layers. It rather produces multi-layer 
networks from heuristic assumptions on the growth process. We are going to present an in-depth
study of network characteristics for both models in comparison in a forthcoming paper.

We will begin by discussing important properties of power grids at various network layers, 
focusing on the distinction between high and very high voltage levels that are typically meshed 
and mid and low voltage levels, that are typically operated as tree networks. We will then discuss 
two network construction algorithms that define a multi-layer power grid in detail and present a selection
of network characteristics from an ensemble analysis.

\section{Multi-layer Networks}

An emerging field of research is the analysis of interactions between more than one interconnected network,
so-called networks of networks \cite{Gao2011a,Bashan2013,Boccaletti2014,DeDomenico2016}. Large developed power grids 
in are themselves networks of networks since they are typically composed of transmission grids interconnecting 
various regional distribution grids. In this regard, we can classify power grids as interconnected or interacting 
networks \cite{Duenas-osorio2004,Donges2011,Gao2011a,Boccaletti2014,VanMieghem2016}.

Among the literature on the network structure of isolated network layers, transmission grids 
are treated most prominently (e.g. \cite{Rosas-Casals2007,Rosas-Casals2009,Schultz2014,Soltan2015})
in contrast to distribution grids are only rarely studied (e.g. \cite{Pagani2011}). 
A review on the research about transmission network topologies is contained in \cite{Pagani2013,Schultz2014}.
There exist growth models for networks of networks generalising preferential attachment \cite{Barabasi1999},
however they are so far restricted to multiplex networks with one-to-one relations between nodes 
in adjacent layers or do not consider a spatial embedding \cite{Criado2012,Nicosia2013,Wang2013}.

The design of growth mechanisms for spatially embedded networks focuses on two central problems, 
namely a) the probability of a new node to appear at a certain position and b) the probability for
a new edge to be added. These ingredients, \emph{growth} and \emph{preferential attachment} have already
been identified to be responsible for the emergence of scale-invariant connectivity distributions as an 
indicator for hierarchical organisation in complex networks \cite{Barabasi1999}.
The first problem is typically addressed by assuming a spatial probability 
distribution of nodes \cite{Schultz2014}, which can also be derived from existing network data \cite{Soltan2015}.
New edges, however, are typically determined by special preferential attachment mechanisms \citep{Duenas-osorio2004}
where the linkage probability is a function of a distance measure between nodes \cite{Schultz2014} and 
various further properties. In \cite{Duenas-osorio2004} a two-layer growth model is proposed, where 
supply and demand ratings of nodes are taken into account and inter-layer edges are only added for high-degree nodes. 

\subsection{Definitions}

Let us define an undirected network $G = (V, E)$, where the node set $V$ and edge set $E$  
comprise the busses and branches of a power grid respectively. Note that $G$ can alternatively
be defined as a weighted network $G=(V, E, W)$ with edge weights $W$ containing line admittances.

We assume there exist two functions $lev(v)$ and $lev(e)$ that encode which level $1\ldots L$ each node $v\in V$ and edge $e\in E$ belongs to. 
The level of an edge will always be the minimum of the two levels of its end nodes.
A layer $G^\ell$ is defined as the subnetwork of $G$ induced by the sets of nodes $v$ and edges $e$ with $lev(v)=\ell$ or 
$lev(e)=\ell$,respectively. $G^\ell$ typically consists of several connected components, i.e. separate regional grids.
Note that $V$ is the disjoint union of all $V^\ell$.

We denote with $G_c^\ell=(V_c^\ell,E_c^\ell)$ the \emph{``cumulative'' subgraph} of $G$
formed by all nodes and edges of level $\ell$ or higher, i.e., with $lev(v)\ge\ell$ or $lev(e)\ge\ell$.
Then, the pair $(\{G^\ell\}_\ell, \{E^{\ell \kay}\}_{\ell,\kay})$ 
of layer subgraphs $G^\ell$ and layer interconnections $E^{\ell \kay}=\{e=(v',v'')\in E | v'\in V^\ell \wedge v'' \in V^\kay\}$
is called a \emph{multi-layer network}. 

The network $G$ is embedded in an external space -- here we use the square $[-1,1]^2$ -- such that
 each node has a location $x(v)\in [-1,1]^2$. We denote by $d_2(v,v') = ||x(v) - x(v')||_2$ 
 the \emph{spatial (Euclidean) distance} between the locations of $v$ and $v'$

Note that all inter-layer edges $E^{\ell \kay}$ connect nodes at the same coordinates, i.e. they correspond to 
transformers modelled as internal edges between busses at different voltage levels. By identifying these
nodes and dropping all inter-layer edges, we can define a \emph{spatial projection network} 
$\mathcal{P}(G) = \tilde{\oplus}_\ell G^\ell$ in analogy to projection networks of multiplex networks 
\cite{Boccaletti2014}. Here, $\tilde{\oplus}$ is the disjoint graph union followed by identifying
nodes with identical coordinates.

\section{Review of Network Topologies}

Power grids have been historically built in several network layers, each of which being 
characterised by a different voltage level. There are extra-high ($>$110\,kV, EHV), high
(36-110\,kV, HV), medium (1-36\,kV, MV) and low voltage ($\leq$1\,kV, LV) networks, 
being interconnected by transformer substations. The extra-high voltage layers are referred to
as \emph{transmission} grids whereas the other layers are considered as \emph{distribution} grids.

Power grids are built in a hierarchical structure in the sense that the coupling scheme
is designed for top-down power-flows, i.e. from production centres in high-voltage network 
layers to consumers attached to low-voltage grids. This is mirrored in a hierarchical control
scheme, where the global balancing of production and consumption is mainly done on transmission
grid level.

Historically, power systems were local small-scale LV (MV) networks \cite{50Hertz, BURN},
mainly in urbanised areas. For instance, the Pearl Street Station in New York City (1882), 
one of the first power stations, served a radius of less then 1~km.
They have been constructed in proximity of load centres to minimise line losses, 
subsequently the LV (MV) network capacity improvement has mainly been driven by the amount of load.
Failures of islanded power plants, fluctuating demand and extreme weather situations posed large
risks to these local distribution networks, while increased line loadings necessitated higher voltage ratings.
Consequently, they have been interconnected using an overlay HV network \cite[p.3]{Lakervi1995}, 
e.g. the UK supergrid built in the 1950's. 
EHV superseded HV in this role and HV networks are primarily used as distribution grids \cite[p.148]{Lakervi1995}.
Transmission grids made it possible to globally balance production and consumption as well as to 
include large-scale power plants into the grid which couldn't be built close to load centres anymore.

\paragraph{Low voltage} networks grow radially from single nodes in the medium voltage network 
proportional to the local load development. As the average load density in cities can exceed $100~MW/km^2$
in contrast to $10~kW/km^2$ in rural areas, there are difference between urban/rural LV grids 
\cite[p.194ff.]{Lakervi1995}, e.g. urban grids tend to be interconnected due to their proximity.
In general, they are operated as radial networks (open loops) \cite[p.192]{Lakervi1995}, i.e. 
with a single in-feed node from MV, where the length of an LV line is usually limited to 500m or 
less \cite[p.12]{Lakervi1995}.

\paragraph{Medium voltage} grids are (as well) operated as radial networks (open loops) \cite[p.17]{Lakervi1995}
as meshed configurations are more complex to control \cite[p.176]{Lakervi1995}. In urban areas, an MV 
grid follows the road network, typically as underground cables \cite[p.181]{Lakervi1995}. Hence,
their network evolution in urban areas probably correlates with road networks \cite{Strano2012}.
In principle, local LV networks could be directly coupled to the HV layer without the need of 
intermediate MV grids. Still, MV grids remain because they have considerably less construction costs
and are more suitable for industrial consumers \cite[p.166]{Lakervi1995}.

\paragraph{Extra-high and High voltage} networks consist of long-range connections with high capacity, 
typically built as a meshed grid \cite[p.15]{Lakervi1995}.
Both HV and EHV usually fulfil the so-called $N-1$-criterion \cite[p.7]{Lakervi1995} which means that
they are required to be resilient towards the failure of a single component.

As a simplifying assumption, we are going to consider three different network layers in the following:
\begin{itemize}
\item low (LV), very local distribution networks
\item middle (MV and HV), the actual distribution backbone in an area
\item high (EHV), connecting different areas over long distances
\end{itemize}

This choice is a trade-off between the observed functional roles or design principles of network layers in power grids
and the modelling complexity needed to reproduce important features. Our model presented in the 
following are, however, not restricted and naturally work for the general case of an arbitrary number 
of network layers. 

Note that the classification of HV grids is somewhat ambiguous. On the one hand, their functional role
relates them to MV grids, while on the other hand, their rather meshed structure is more similar to EHV grids.
We use the categorisation given above, however, this distinction might differ strongly across 
real power grids and should be evaluated on a case-by-case basis.

\section{Model Description}

\subsection{Recapitulating the Mono-layer Model \cite{Schultz2014}} \label{sec:basemodel}
In a previously published random growth model \cite{Schultz2014} for synthetic 
infrastructure networks, we describe a heuristic growth process to create statistically
suitable network topologies. It has been especially tuned to high-voltage transmission networks and
consists of two growth phases which we are going to recapitulate in the following.
For details of the implementation, we refer to \cite{Schultz2014}.

In the first phase (initialisation), an initial number of nodes $n_0$ is connected using 
a minimum spanning tree, minimising the overall edge length. Furthermore, a number of 
additional edges $m=n_0 (1-s)(p+q)$ are chosen to connect the node pairs for which a 
cost function 
\begin{align}
f(v,v')=\frac{(d_G(v, v') + 1)^r}{d_2(x(v), x(v'))}
\end{align}
is maximal. By $d_G(v,v')$ we denote the minimum length of any path from $v$ to $v'$ 
in the network $G$.  Here, a control parameter $r$ determines the value of additional redundancy 
for adding a new edge $i$--$j$ between two nodes at positions $x(v)$ and $x(v')$.

The second phase (growth) consecutively adds new nodes to the existing network using 
an attachment rule that can be adjusted with three parameters $p$, $q$ and $s$. 
They correspond to a probability $p$ of connecting a new node with a second (redundant) edge, 
a probability $q$ of reinforcing the network by connecting a pair of existing nodes and a 
probability $s$ to split an existing edge by adding the new node inbetween.

In summary, a set of five independent parameters $\{n_0,p,q,r,s\}$ along with the desired
network size $n$ fully determines the model output. 
Additionally, the network characteristics also depend on the choice of
node locations. They can be drawn at random from a given spatial distribution, determined
from data, or both. For instance, a Gaussian mixture model has been applied to locational 
data from US power grids \cite{Soltan2015} to obtain node location densities.

We propose two different extensions of this base model. On the one hand, we consider
a static construction by 'gluing' together network layers according to a specified 
layer interconnection structure. The inter-layer networks are derived from the base model
with individually adjusted parameters. On the other hand, we allow the parallel growth
of network layers at a different rate. In this way, we obtain a multi-layer network 
without predefined layer interconnections.

\subsection{Static Glue Stick Construction (\GSC)}

If network parameters are initially known (node positions, branching values etc.), the base
model can be used to construct $G$ from its layers $G^\ell$ respectively each layer $G^\ell$
from a set of connected components $C_i^\ell$ using an individual set of base model parameters 
$\{n^\ell, n_0^\ell, p^\ell,q^\ell,r^\ell,s^\ell\}$.
Here, we are going to assume homogeneous model parameters for all components of a layer $\ell$ but
in general it might be useful to consider certain parameter distributions.
Different layers are connected via transformers which are represented as internal edges 
between a high-voltage and low-voltage bus. 

The \GSC model can be set up as a bottom-up algorithm starting from the lowest layer and 
consecutively adding higher ones or reversed (top-down), depending on the model scenario.
The resulting multi-layer networks, however, do not differ qualitatively.

\begin{figure*}[!ht]
    \centering
    \begin{subfigure}[t]{0.5\textwidth}
        \centering
        \includegraphics[width=\textwidth]{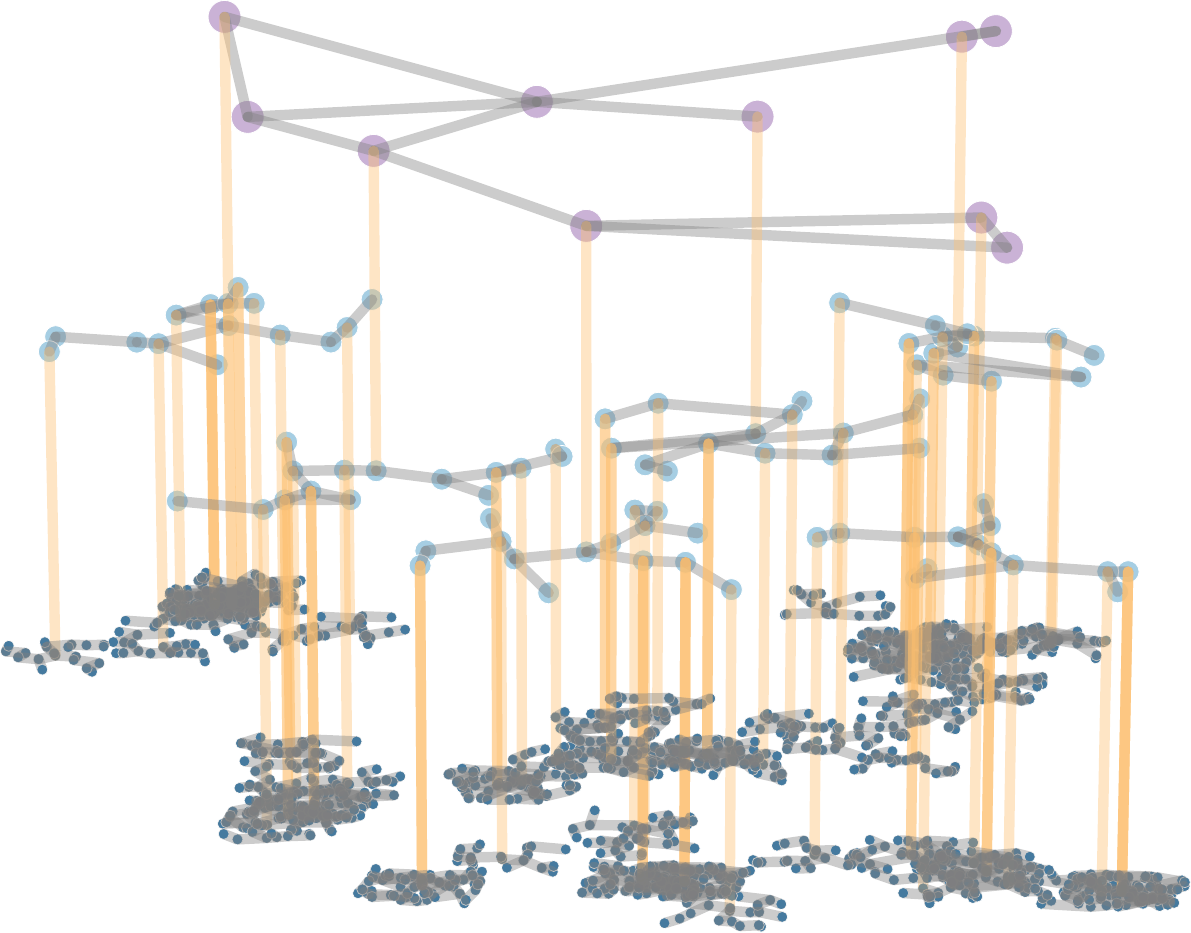}
        \caption{}
    \end{subfigure}%
    ~ 
    \begin{subfigure}[t]{0.5\textwidth}
        \centering
        \includegraphics[width=\textwidth]{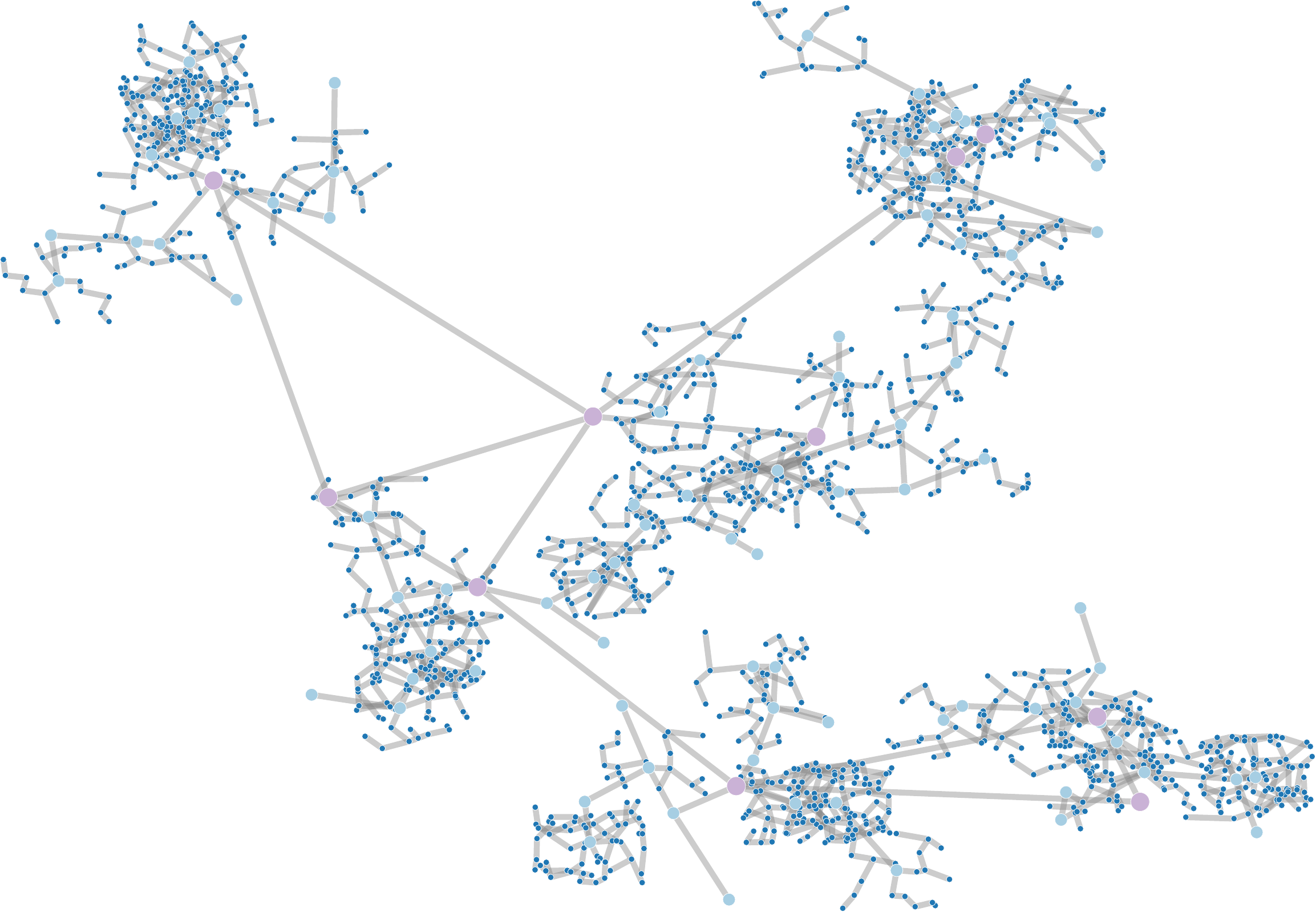}
        \caption{}
    \end{subfigure}
    \caption{(a) Schematic representation of a multi-layer network $G$ created with the \GSC algorithm.
It consists of three layers (i.e. high/middle/low) which are interconnected by transformers,
here visualised as vertical (orange) edges. The layers are separated according to their voltage 
rating. The model parameters are listed in the text.
(b) Spatial projection network $\mathcal{P}(G)$ of the multi-layer network depicted in (a).}
	\label{fig:gluestick}
\end{figure*}

Fig.\,\ref{fig:gluestick}(a) pictures the output of this algorithm using the parameters $n=(10, 15, 12)$, 
$n_0=(8, 14, 11)$, $p=(.2, .1 ,0)$, $q=(.3, 0, 0)$, $r=(.33, .33, .33)$, $s=(.1, .1, .1)$. 
Node colour corresponds to the different network layers, whereas edge colours differentiate between 
inter- and intra-layer edges. While the $G^{high}$ consists of a single connected component, 
$G^{middle}$ and $G^{low}$ consist of 6 respectively 85 components. In Fig.\,\ref{fig:gluestick}(b),
we further show the corresponding spatial projection network.

\subsection{Interconnected Layer Growth (\ILG)}

We now present a more generic network generation algorithm that is relatively simple but still quite flexible 
and may be used to generate synthetic infrastructure networks representing 
power grids, transportation networks, or communication networks 
in which nodes and edges may reside on any number of levels.
It is basically a multi-layer version of the algorithm described at the beginning of this section \cite{Schultz2014}
with some small changes in the trade-off function used for the placement of redundant edges
(now using spatial length weighted network distance and introducing preference for target nodes in more densely populated areas).

The algorithm forms an undirected network $G=(V,E)$ along with node locations $x(v)\in [-1,1]^2$ for each $v\in V$, 
We denote by $d_{G_c^\ell}(v,v')$ the minimum weighted length of any path from $v$ to $v'$ in $G_c^\ell$ 
where an edge from $v''$ to $v'''$ has weight $d_2(v'',v''')$.
In addition, we define a ``node density'' measure by putting
\begin{align}
	d(v) = \sum_{v'\in V_c^{lev(v)}\setminus\{v\}} d_2(v,v')^{-2},
\end{align}
which will be used as a simple proxy for the ``user population'' per area ``around'' $x(v)$.

Note that we restrict the sum to nodes of at least the same level as $v$.
In this way, the construction of each level will be independent of the lower levels,
which we consider a useful property since it allows one to use our model with different values of $ L$
and get similar topologies for the highest level(s) no matter how many lower levels one chooses to model in addition.

To govern the placement of redundant edges, we define a ``trade-off'' function 
\begin{align}
	f^\ell(v,v') = \frac{(d_{G_c^\ell}(v,v') + d_2(v,v'))^{r^\ell} d(v')^{u^\ell}}{d_2(v,v')}.
\end{align}
The idea is that building a redundant edge from $v$ to $v'$ has 
costs proportional to spatial length $d_2(v,v')$
and benefits resulting from 
(i) forming a redundant cycle of length $d_{G_c^\ell}(v,v') + d_2(v,v')$
and (ii) simplifying access from $v$ to a ``user population'' proportional to $d(v')$.
The trade-off function thus values the prospective line by combining the two types of benefits 
into a Cobb-Douglas type ''utility'' assessment $(d_{G_c^\ell}(v,v') + d_2(v,v'))^r d(v')^u$,
and then computing utility per unit cost.
In this, $r^\ell,u^\ell\ge 0$ are parameters governing the importance of the benefits.
Our usage of $d(v')^{u^\ell}$ was inspired by 
a similar term used in \cite{Soltan2015} for calculating line placement probabilities,
which was however based on a different definition of density that was less fast to calculate and update during network growth.

To place a new node $v$ into some level $\ell$, the algorithm will make use of the following ``placement step''
\begin{itemize}
\item[(P)]	Draw $y$ uniformly at random from the square $[-1,1]^2$,
			draw $v'$ uniformly at random from $V_c^\ell\setminus\{v\}$. 
			Then either put $x(v) = \alpha^\ell y + (1-\alpha^\ell) x(v')$ (with probability $\gamma^\ell$)
			or put $x(v) = \beta^\ell y + (1-\beta^\ell) x(v')$ (with probability $1 - \gamma^\ell$).
\end{itemize}

Now, for each phase $\phi = 1\ldots L$, the algorithm performs the following steps:
\begin{itemize}
\item 	Introduction and initialisation of a new level $\ell=\phi$:
	\begin{itemize}
	\item[I1]	Add $n_0^\ell$ many nodes $v_i^\ell$ to $V$, with $lev(v_i^\ell)=\ell$ and
				random locations $x(v_i^\ell)$ as follows.
		\begin{itemize}
		\item[I1.1]	Draw $x(v_1^\ell)$ uniformly at random from the square $[-1,1]^2$.	
		\item[I1.2] For $i=2\ldots n_0^\ell$, perform step (P) for node $v=v_i^\ell$. 
		\end{itemize}
	\item[I2]	Find the minimum spanning tree (w.r.t.\ Euclidean distance) of these $n_0^\ell$ many locations 
				and add all its edges $e$ to $E$, putting $lev(e)=\ell$.
	\item[I3]	Add $m = \lfloor n_0^\ell(1-s^\ell)(p^\ell+q^\ell) \rfloor$ many redundant edges to level $\ell$ as follows. 
				For $a = 1\ldots m$, draw a $v\in V^\ell$ uniformly at random,
				find that $v'\in V^\ell\setminus\{v\}$ that has no edge to $v$ yet for which $f^\ell(v,v')$ is maximal,
				add a new edge $e$ from $v$ to $v'$ to $E$, and put $lev(e)=\ell$. 
	\item[I4]	If $\ell>1$, connect the previous level to the new one by finding the node $v\in V^{\ell-1}$
				that minimises $d_2(v_1^\ell,v)$, adding a new edge $e$ from $v_1^\ell$ to $v$ to $E$, and putting $lev(e)=\ell-1$.
	\end{itemize}
\item	Simultaneous growth of all already existing levels $\ell = 1\ldots\phi$.
		For each $\ell = 1\ldots\phi$, 
		let $U^\ell$ be a set of $n_\phi^\ell$ many new nodes $v$ to be added with $lev(v)=\ell$,
		and let $U$ be the union of all these $U^\ell$.
		For each $v\in U$, drawn uniformly at random without replacement, do the following:
	\begin{itemize}
    \item[G0]	Add $v$ to $V$ and let $\ell=lev(v)$. 
    			With probabilities $1-s^\ell$ and $s^\ell$, perform either steps G1---G4 or step G5 below, respectively.
    \item[G1] 	Perform step (P) to determine $x(v)$.
    \item[G2]	Find that node $v'\in V_c^\ell\setminus\{v\}$ for which $d_2(v,v')$ is minimal,
    			add a new edge $e$ from $v$ to $v'$ to $E$, and put $lev(e)=\ell$.
	\item[G3]	Draw a number $k\ge 0$ from the geometric distribution with mean $p^\ell$
				and repeat the following $k$ times:
				find that node $v'\in V_c^\ell\setminus\{v\}$ that has no edge to $v$ yet for which $f^\ell(v,v')$ is maximal,
				add a new edge $e$ from $v$ to $v'$ to $E$, and put $lev(e)=\ell$. 
	\item[G4]	Draw a number $k\ge 0$ from the geometric distribution with mean $q^\ell$
				and repeat the following $k$ times:
				draw a node $v''\in V^\ell$ uniformly at random,
				find that node $v'\in V_c^\ell\setminus\{v''\}$ that has no edge to $v''$ yet for which $f^\ell(v'',v')$ is maximal,
				add a new edge $e$ from $v''$ to $v'$ to $E$, and put $lev(e)=\ell$. 
	\item[G5]	Select an edge $e\in E^\ell$ uniformly at random, 
				let $v',v''$ be its end nodes,
				draw $a\in[0,1]$ uniformly at random,
				let $x(v) = ax(v') + (1-a)x(v'')$, 
				remove $e$ from $E$ and add two new edges $e',e''$ with $lev(e')=lev(e'')=\ell$ to $E$,
				one from $v'$ to $v$, the other from $v$ to $v''$.
	\end{itemize}
\end{itemize}

The parameters are the following: 
\begin{itemize}
\item $L\ge 1$: number of levels. 
\item $n_0^\ell\ge 1$: number of initial nodes of level $\ell$ at its introduction.
\item $n_\phi^\ell\ge 0$: number of additional nodes of level $\ell$ grown in phase $\phi\ge\ell$.
\item $\alpha^\ell,\beta^\ell,\gamma^\ell\in[0,1]$: node location distribution parameters governing the amount of spatial clustering.
\item $p^\ell$: expected number of redundant edges each new node gets immediately.
\item $q^\ell$: expected number of additional redundant edges added to random nodes at each growth step.
\item $r^\ell,u^\ell\ge 0$: importance of redundant edge benefits.
\item $s^\ell$: rate of edge splittings.  
\end{itemize}

\begin{figure*}
	\begin{centering}\noindent
		\includegraphics[width=\textwidth]{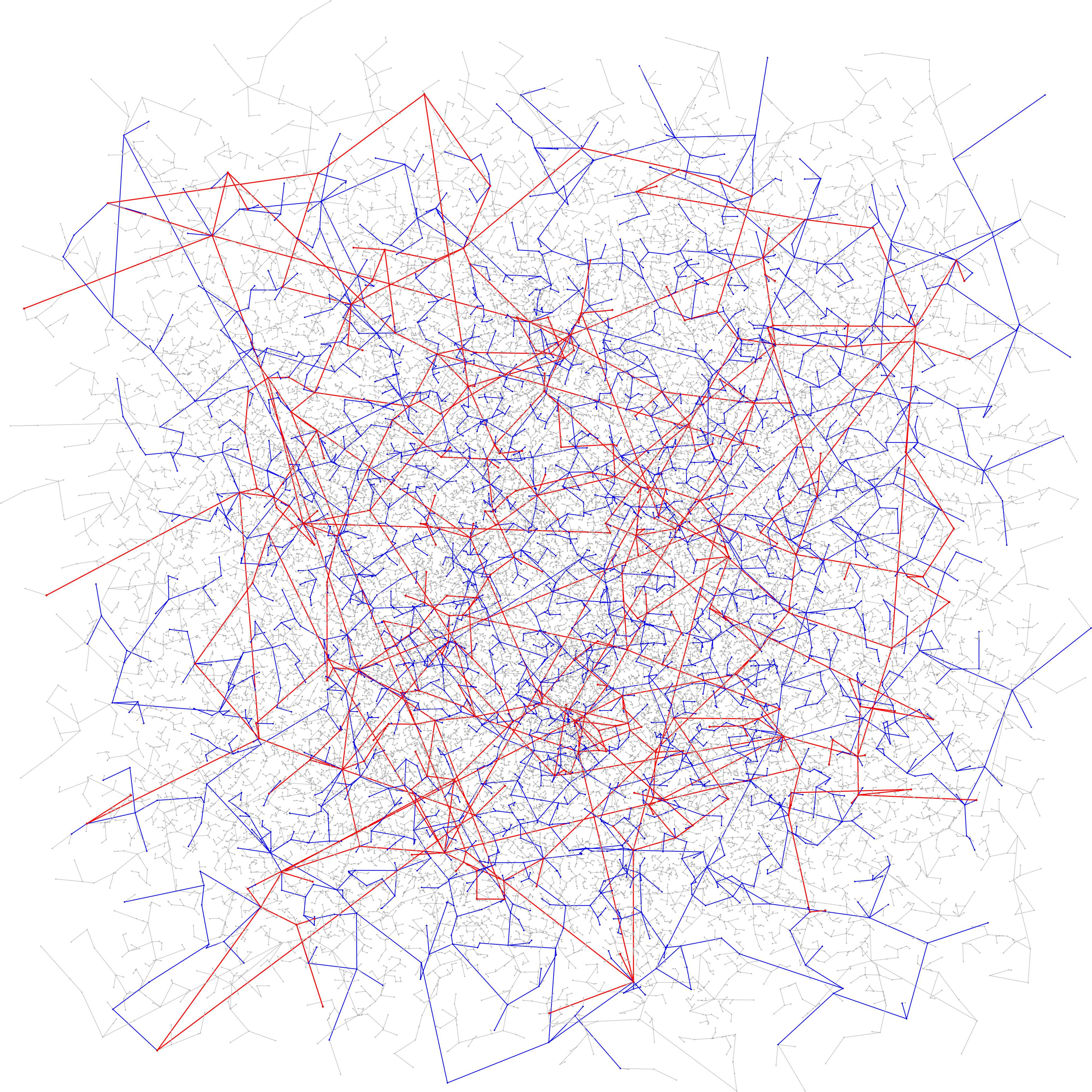}
	\end{centering}
	\caption{
		\label{fig:jobst1} 
		Example of a three-level network with 50,000 nodes and moderate spatial clustering,
		grown in three phases:
		(i) a minimal spanning tree (MST) of 500 nodes in the lower level (grey);
		(ii) a MST of 250 nodes in the middle level (blue);
		(iii) a MST of 100 nodes in the upper level (red), followed by growing and connecting 250 red, 2400 blue and 46,500 grey nodes.
		Pictured is the spatial projection network $\mathcal{P}(G)$.
		The lower level contains no redundant edges and is thus a union of disjoint trees connected to the higher levels.
		Further parameters (see text):
		$\alpha^\ell\equiv .05$, $\beta^\ell\equiv .5$, $\gamma^\ell\equiv .5$,
		$p=(.3,.1,0)$,
		$q=(.075,.075,0)$,
		$r=(1.5,.75,0)$,
		$u=(.1,.05,0)$,
		$s=(0,.05,.2)$.
	}

\end{figure*}

Fig.\,\ref{fig:jobst1} shows an example with $L=3$ and a total of 50,000 nodes, generated in under one hour on an Intel i7-6600U CPU with under one GB of memory usage.

For an efficient implementation of the algorithm,
one can keep in memory a list of node densities $d(v)$ 
for just those $v\in V$ with $lev(v)\ge$the minimal $\ell$ for which $u^\ell>0$.
Then $d(v)$ with $lev(v)=\ell$ is easily updated whenever a node $v'$ with $lev(v)\ge\ell$ is added
by simply adding $d_2(v,v')^{-2}$ to $d(v)$.

Storing a full spatial distance matrix can be avoided by using efficient spatial data structures such as R${}^+$-trees.
If $p,q$ are relatively small (as in power grids), $f$ needs to be evaluated rarely,
thus storing and updating a full network distance matrix would be inefficient.
Hence, the algorithm can also be implemented efficiently with low memory requirements. 

\section{Ensemble Network Characteristics}

In the following section, we give an impression of network characteristic for
multi-layer power grids created with both models.

A number of different network metrics has been proposed to characterise
power grids (e.g. \cite{Schultz2014,Soltan2015}. Commonly used are 
degree and edge length distributions. 

We consider an ensemble of 50 multi-layer networks with $L=3$ layers, constructed with each model.
The \GSC model parameters are $n=(10, 15, 12)$, 
$n_0=(8, 14, 11)$, $p=(.2, .1 ,0)$, $q=(.3, 0, 0)$, $r=(.33, .33, .33)$, $s=(.1, .1, .1)$.
This choice results in a total number of 1690 nodes.

The \ILG parameters are chosen to be 
$\alpha^\ell\equiv .05$, $\beta^\ell\equiv .5$, $\gamma^\ell\equiv .5$,
$n_\phi=(985, 5, 10)$
$n_0=(2, 5, 10)$
		$p=(.3,.1,0)$,
		$q=(.075,.075,0)$,
		$r=(1.5,.75,0)$,
		$u=(.1,.05,0)$,
		$s=(0,.05,.2)$.
This choice results in a total number of 1000 nodes.

\begin{figure*}[h!]
    \centering
    \begin{subfigure}[t]{0.5\textwidth}
        \centering
        \includegraphics[width=\textwidth]{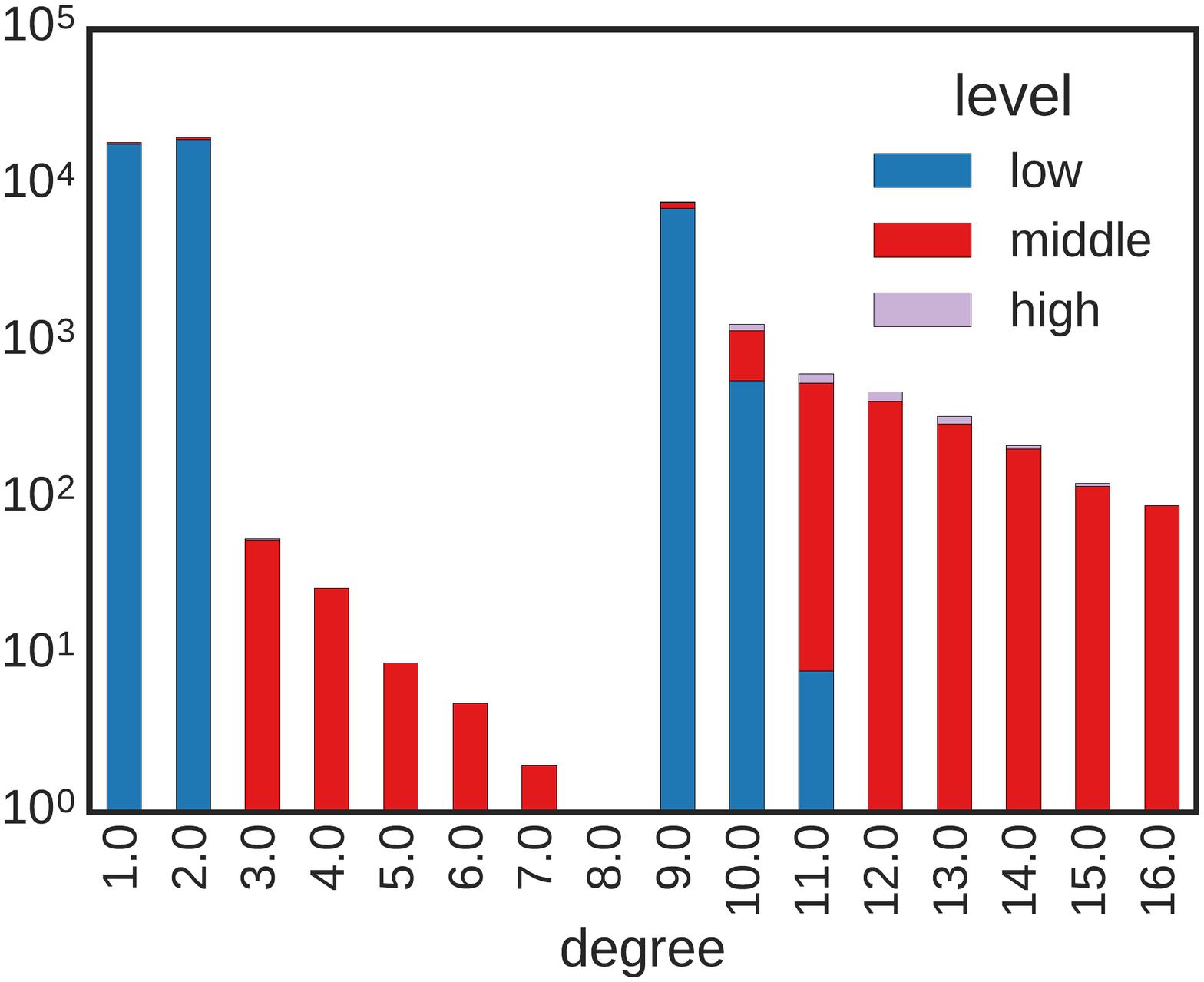}
        \caption{}
    \end{subfigure}%
    ~ 
    \begin{subfigure}[t]{0.5\textwidth}
        \centering
        \includegraphics[width=\textwidth]{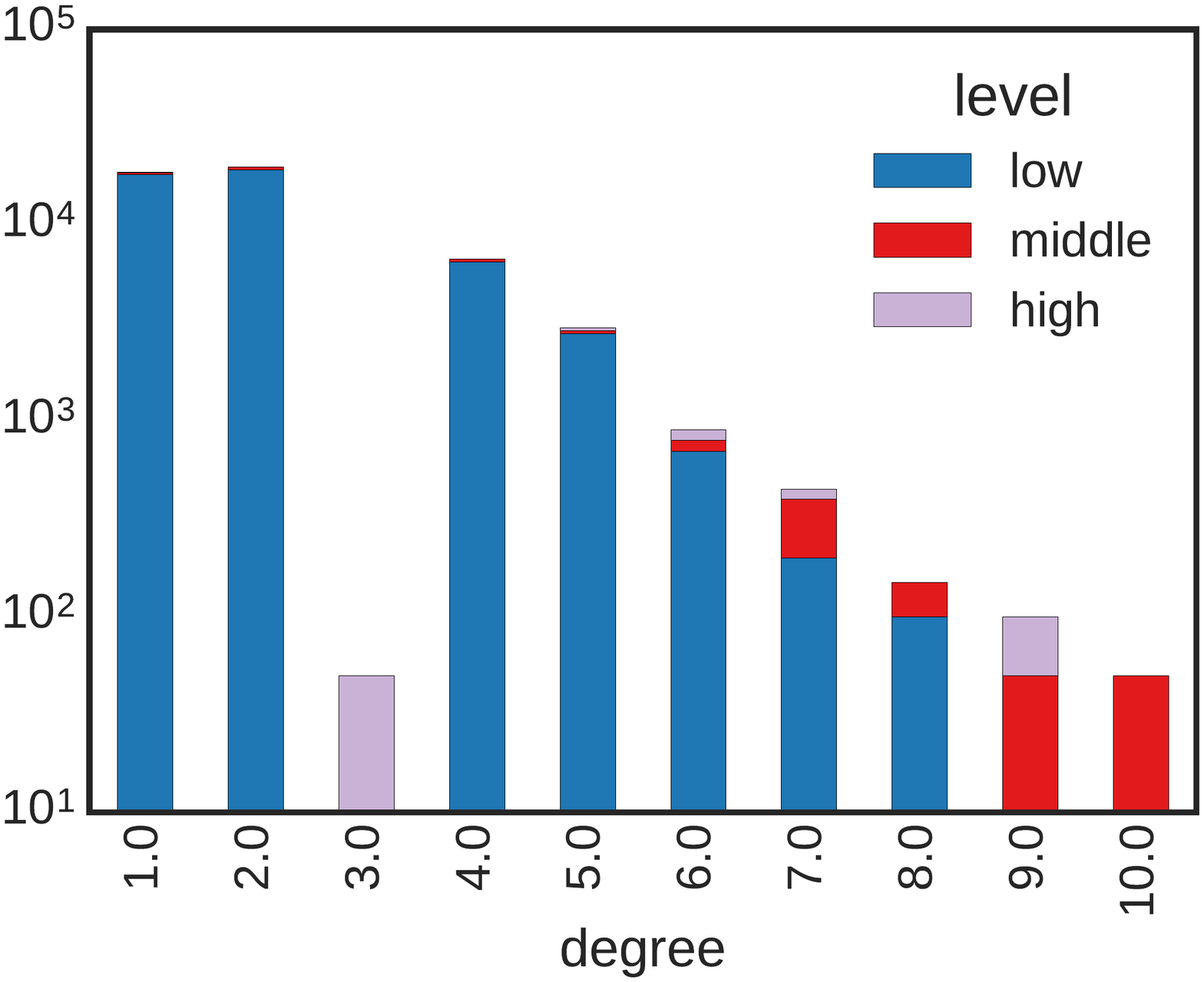}
        \caption{}
    \end{subfigure}
    \caption{Degree distribution by level from ensembles of 50 multi-layer power grids created with (a) the \GSC (b) the \ILG model.
     }
	\label{fig:degree}
\end{figure*}

In Fig.\,\ref{fig:degree}(a) and (b), it can be seen that both models yield 
networks with roughly geometric  degree distributions of $\mathcal{P}(G)$. 
This result shows  that higher-level nodes tend to have larger degrees. 
Interestingly, for our choice of parameters, we observe gaps in the distribution. 
This is a result of the projection where nodes $v, v'$ with the same coordinates
are identified as a new node $v''$ whose degree becomes $k(v'') = k(v) + k(v') - 2$.

\begin{figure*}[h!]
    \centering
    \begin{subfigure}[t]{0.5\textwidth}
        \centering
        \includegraphics[width=\textwidth]{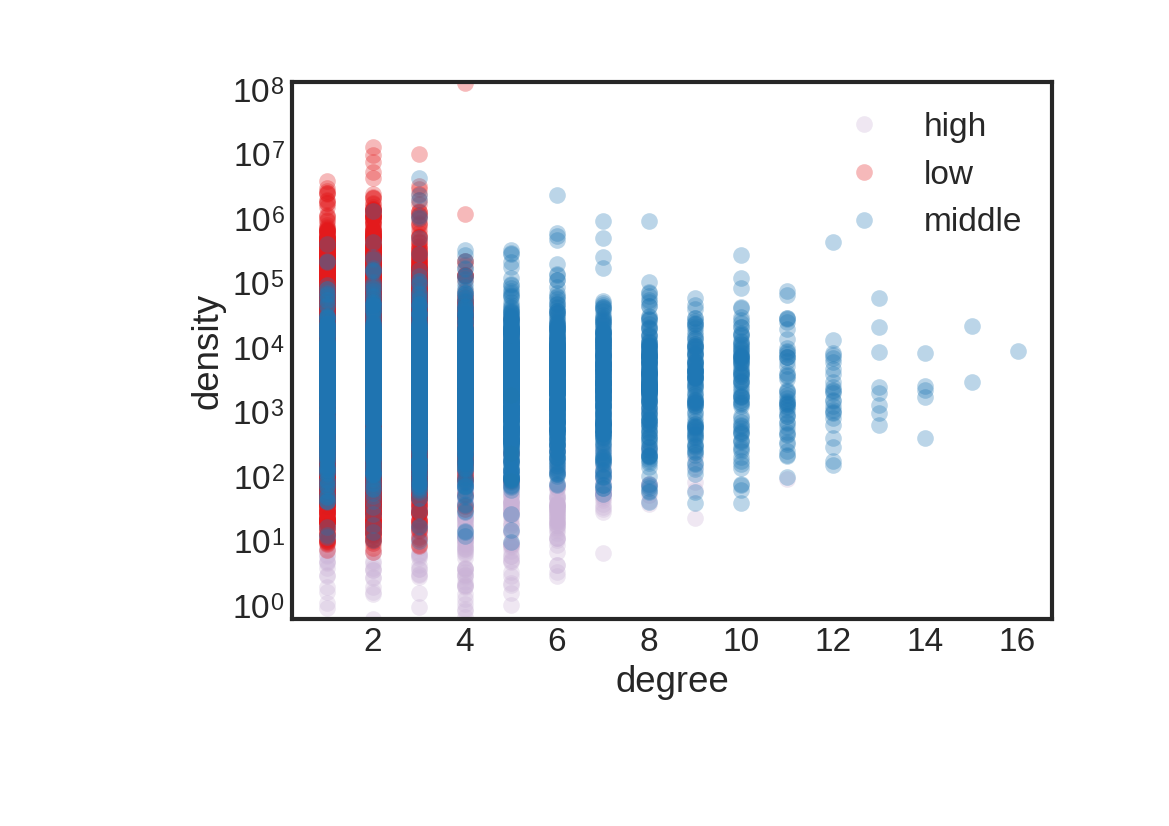}
        \caption{}
    \end{subfigure}%
    ~ 
    \begin{subfigure}[t]{0.5\textwidth}
        \centering
        \includegraphics[width=\textwidth]{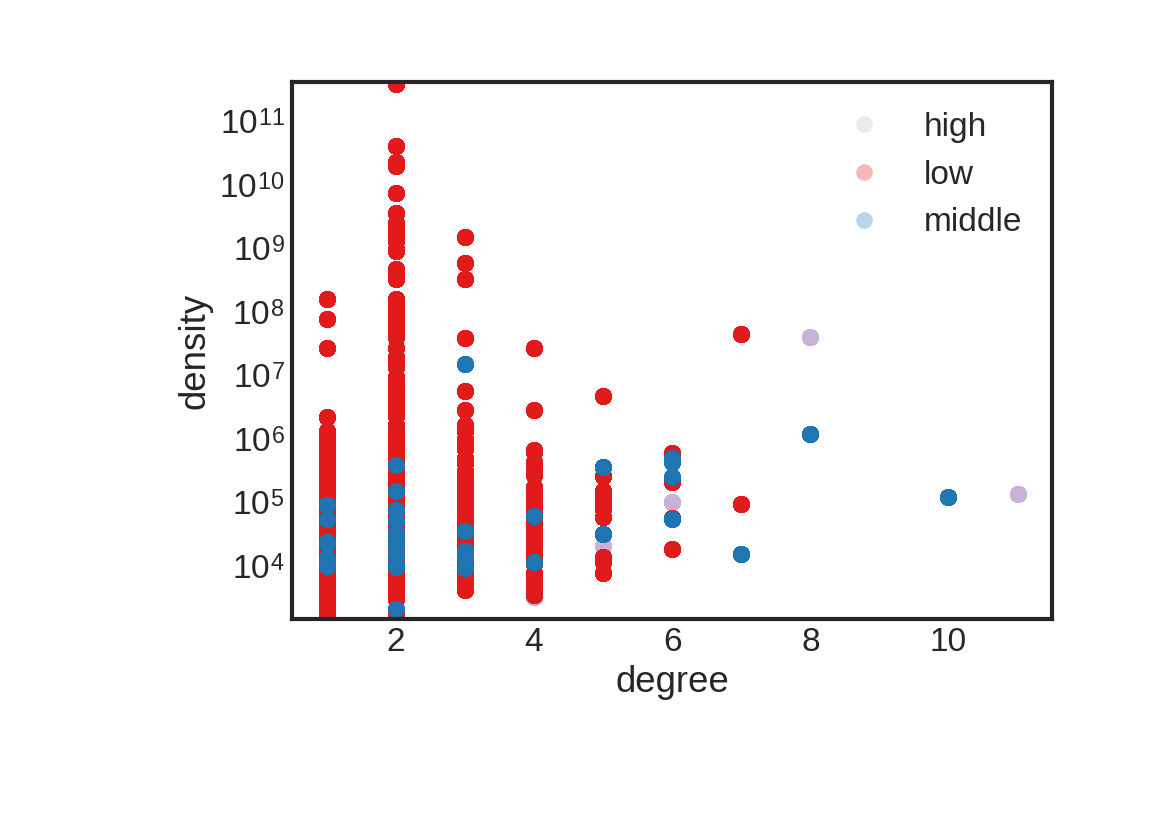}
        \caption{}
    \end{subfigure}
    \caption{Logarithmic edge length distribution by level from ensembles of 50 multi-layer power grids created with the (a) \GSC (b) \ILG model. }
	\label{fig:density_degree}
\end{figure*}

Plotting node density vs. degree of the network (Fig.\,\ref{fig:density_degree}) by level,
		shows a slight positive correlation for the higher-degree nodes, 
		slightly similar to what was reported in \cite{Soltan2015}.
		(Some noise was added to degrees to get a better view of the distribution.
		
\begin{figure*}[h!]
    \centering
    \begin{subfigure}[t]{0.5\textwidth}
        \centering
        \includegraphics[width=\textwidth]{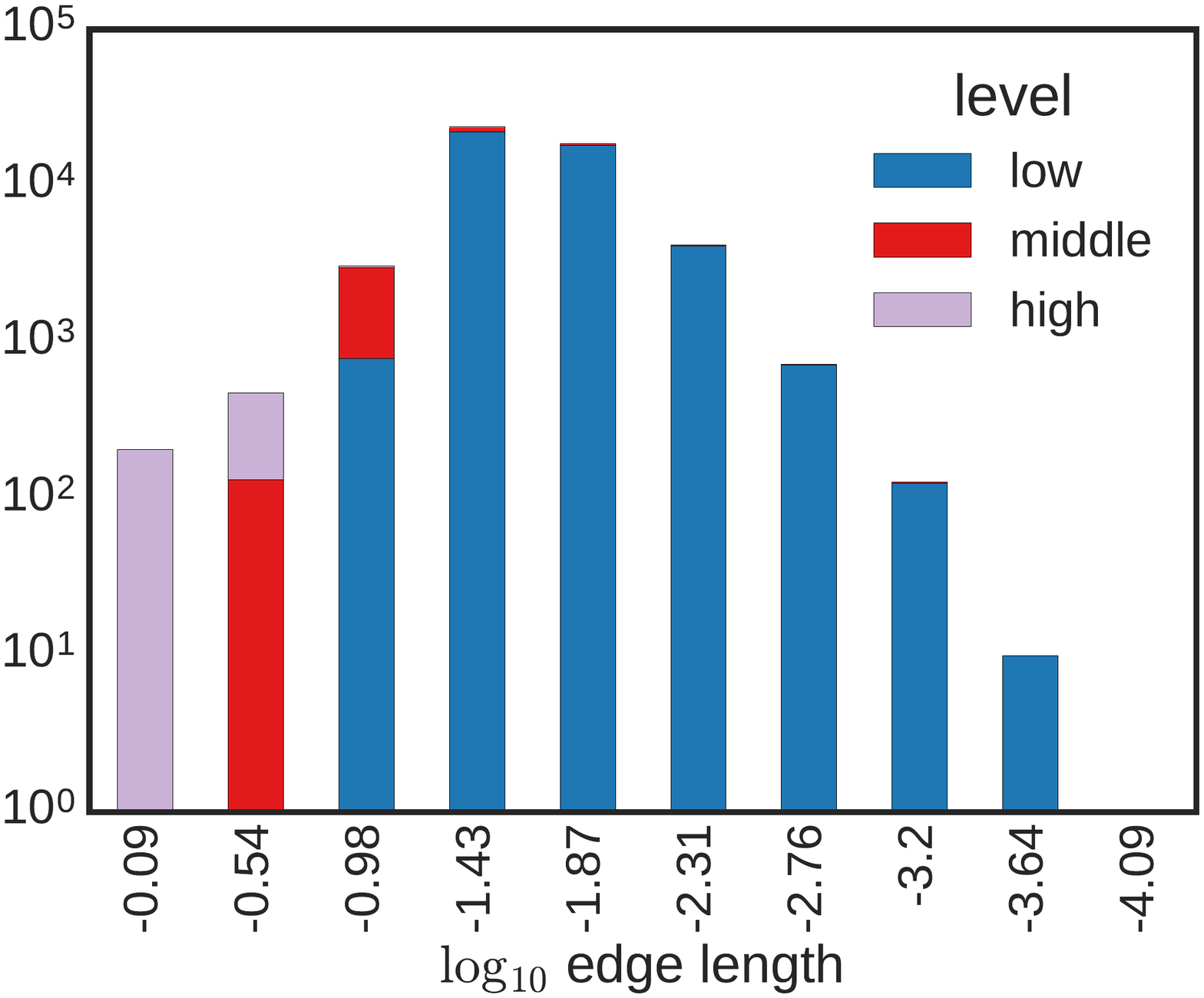}
        \caption{}
    \end{subfigure}%
    ~ 
    \begin{subfigure}[t]{0.5\textwidth}
        \centering
        \includegraphics[width=\textwidth]{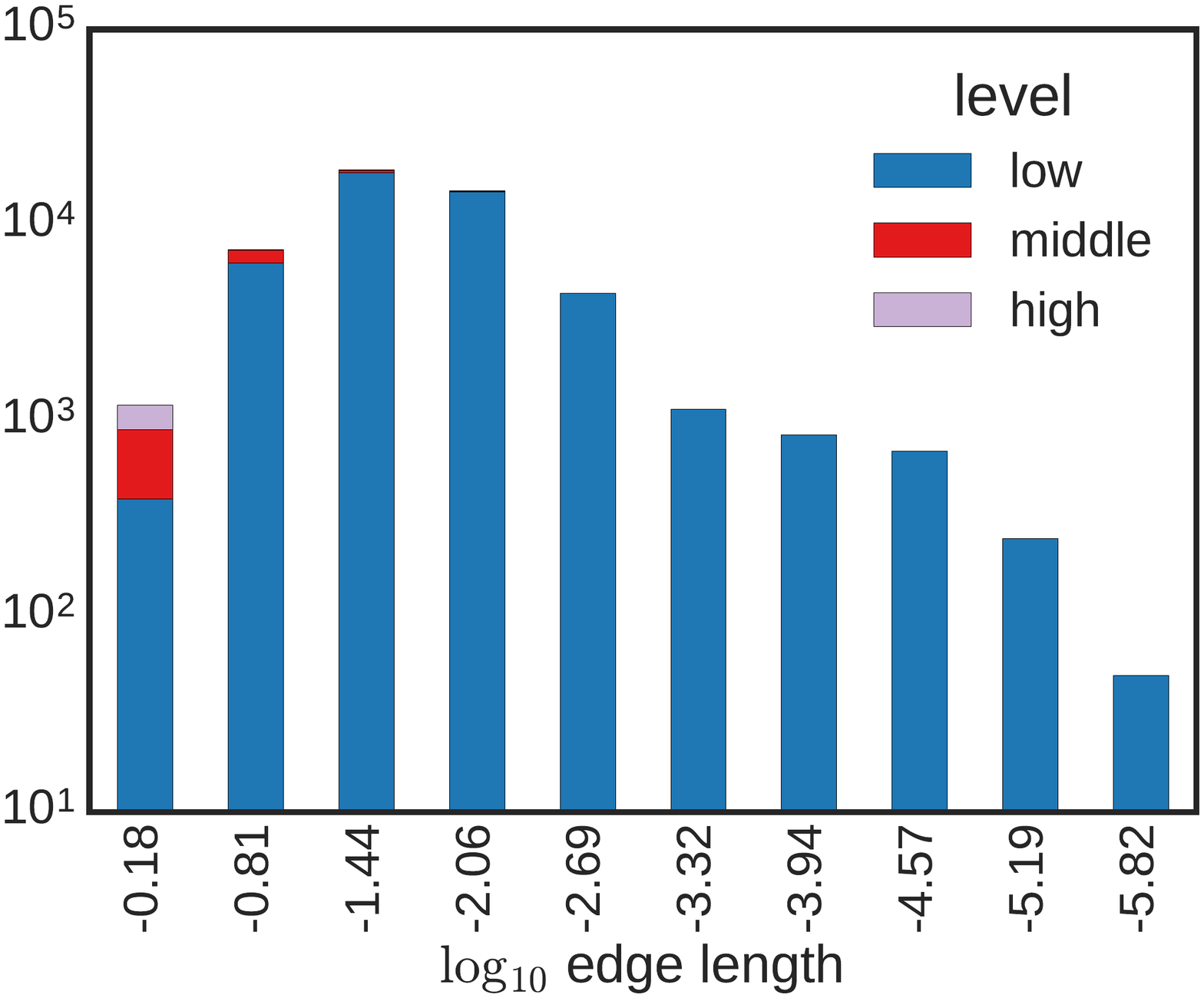}
        \caption{}
    \end{subfigure}
    \caption{Logarithmic edge length distribution by level from ensembles of 50 multi-layer power grids created with the (a) \GSC (b) \ILG model. }
	\label{fig:loglength}
\end{figure*}

The logarithmic edge length distribution is depicted in Fig.\,\ref{fig:loglength}.
The \GSC model yields a two-sided approximately exponential distribution and shows
a strong separation of scales between the layers compared to \ILG networks. 
For the large network depicted in Fig.\,\ref{fig:jobst1}, we find a two-sided approximately exponential distribution
with slopes of roughly $0.9$ (left) and $-1.8$ (right), 
		very similar to what was reported in \cite{Soltan2015} 
		for the real-world North-American and Mexican power grid,
		where the slopes are just a little lower (roughly $0.7$ and $-1.6$).

\section*{Acknowledgement}

The authors acknowledge gratefully the support of BMBF, CoNDyNet, FK. 03SF0472A. 

\bibliographystyle{unsrtnat}
\bibliography{references}

\end{document}